\documentclass[floatfix,twocolumn,aps,showpacs]{revtex4}
\usepackage{amssymb}
\usepackage{graphics,epsf}
\newcommand{\be}{\begin{equation}}
\newcommand{\ee}{\end{equation}}
\newcommand{\edo}{\eta \downarrow 0}
\begin{document}
\title{Cauchy horizon stability in self-similar collapse: scalar radiation.}
\author{Brien C Nolan\footnote{Electronic address:
brien.nolan@dcu.ie} and Thomas J Waters\footnote{Electronic
address: thomas.waters2@mail.dcu.ie}} \affiliation{School of
Mathematical Sciences, Dublin City University, Glasnevin, Dublin
9, Ireland.}
%\date{\today}
\begin{abstract}
The stability of the Cauchy horizon in spherically symmetric
self-similar collapse is studied by determining the flux of scalar
radiation impinging on the horizon. This flux is found to be
finite.
\end{abstract}
 \maketitle
%_____________________________________________________________________
%_____________________________________________________________________
\newtheorem{assume}{Assumption}
\newtheorem{theorem}{Theorem}
\newtheorem{prop}{Proposition}
\newtheorem{corr}{Corollary}
\newtheorem{lemma}{Lemma}
\newtheorem{definition}{Definition}
\newcommand{\so}{{\cal{O}}}
\newcommand{\ch}{{\cal{H}}}
\newcommand{\cf}{{\cal{F}}}
\newcommand{\cm}{{\cal{M}}}
\newcommand{\pnc}{{\cal{N}}}
%_______________________________________________________________________
%______________________________________________________________________
\section{Introduction}
Perhaps the richest source of examples of space-times admitting
naked singularities is the class of spherically symmetric
self-similar space-times. There is an extensive literature on the
topic; the recent review of self-similarity in general relativity
by Carr and Coley \cite{carr} provides a suitable bibliography. Of
particular note in this class are the perfect fluid solutions
studied by Ori and Piran \cite{OP}, the massless scalar field
solutions studied by Christodoulou \cite{christo1} and by Brady
\cite{brady} and the $SU(2)$ sigma model solutions studied by
Bizon and Wasserman \cite{bizon}. We mention these because (i) the
matter model has particular interest, for either physical or
mathematical reasons and (ii) these self-similar solutions are of
interest in studies of critical phenomenon \cite{gundlach}. More
generally, self-similar solutions admitting naked singularities
are of interest because of what they may tell us about cosmic
censorship. Intriguingly, the evidence is not all in one
direction. Recent work has indicated the stability of perfect
fluids admitting naked singularities in the class of perfect fluid
space-times \cite{harada}, while for the case of the massless
scalar field, generic spherical perturbations of self-similar
initial data which correspond to naked singularities will lead to
censored singularities \cite{christo2}. Also, within the class of
self-similar spherically symmetric space-times, the sectors
corresponding to censored and to naked singularities are both
topologically stable \cite{nolan}.

With these results in mind, the aim here is to begin a
comprehensive study of the stability of Cauchy horizons in
self-similar collapse. In the case of charged rotating black
holes, the instability of the Cauchy (or inner) horizon has been
firmly established (see \cite{brady2} for a review). This
instability is in one way easily understood; an observer crossing
the inner horizon views the entire history of the external
universe in a finite amount of proper time, and so time-dependent
perturbations of the exterior suffer an infinite blue-shift on
crossing the horizon. This instability mechanism which can be
``read off'' the conformal diagram, does not have a counterpart in
self-similar collapse which leads to globally naked singularities
(see Figures 1 and 2). At best, one can speculate that the
curvature at the regular center which diverges in the limit as the
scaling origin is approached makes itself felt by perturbations
approaching the Cauchy horizon. This is by no means convincing,
and so a rigorous analysis is required. We begin this analysis
here by studying the propagation of scalar radiation in a fixed
background (spherically symmetric, self-similar) space-time which
admits a Cauchy horizon.

%%Figure One
\begin{figure}[cpic]
\centerline{\epsfxsize=6cm \epsfbox{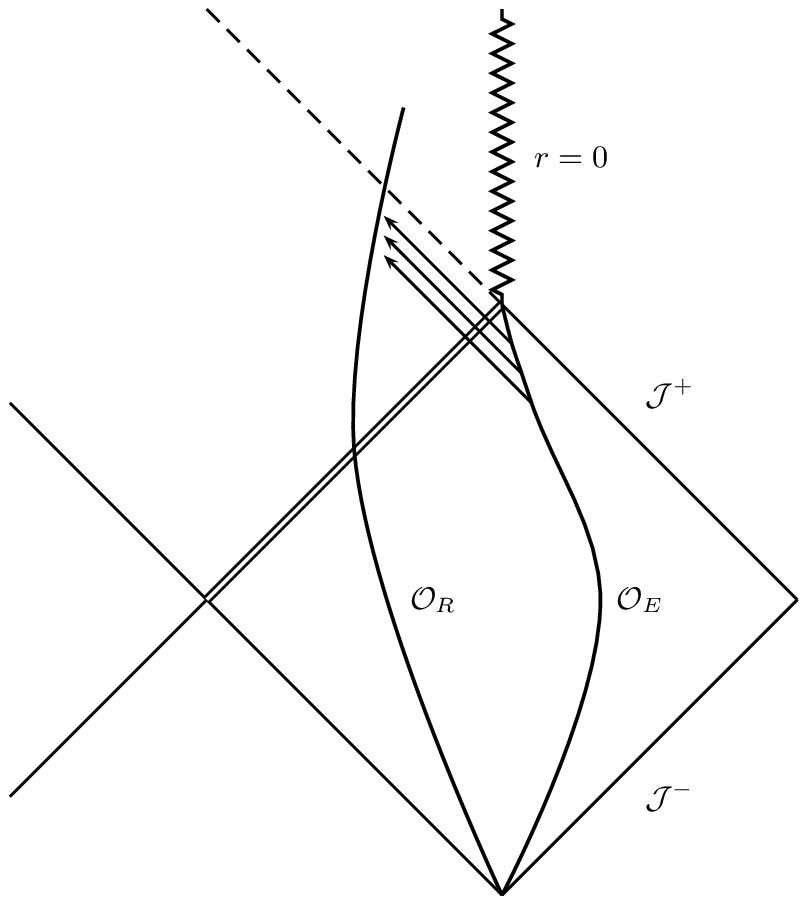}} \caption{A
portion of the conformal diagram of the maximally extended
Reissner-Nordstr\"om space-time. The observer $\cal{O}_R$ falls
through the event horizon (double line) and into the black hole.
On crossing the Cauchy horizon (dashed) into a new asymptotically
flat region, ${\cal{O}}_R$ receives in finite time all the
radiation emitted by ${\cal{O}}_E$ during its infinite history.}
\end{figure}

%Figure 3
\begin{figure}[cpic]
\centerline{\epsfxsize=6cm \epsfbox{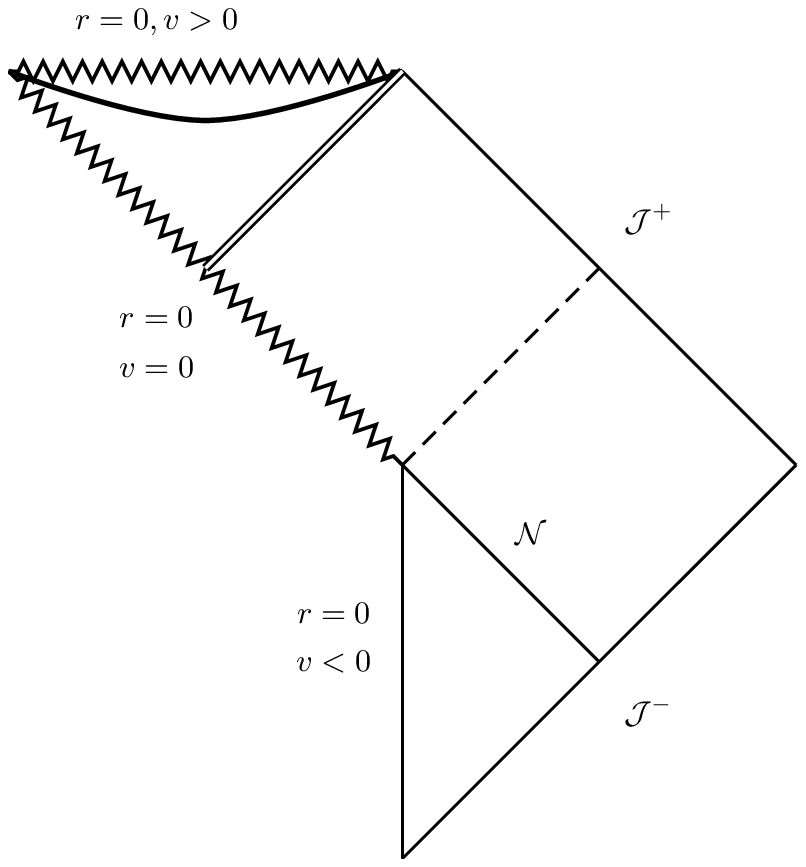}}
\caption{Conformal diagram for an example of a self-similar
space-time admitting a globally naked singularity. We use the
advanced Bondi co-ordinates $v$ and $r$ described in Section 2.
The Cauchy horizon is shown dashed, the event horizon as a double
line and the apparent horizon as a bold curve. $\pnc$ is the past
null cone of the scaling origin. Other structures can arise; there
may be no apparent or event horizon; the censored portion of the
singularity may be null; the naked portion of the singularity may
be time-like. There is evidence that the naked singularity is
generically globally naked. See \cite{nolan} for details. In every
case for which the singularity is naked, the conformal diagram
fails to display an obvious mechanism by which the Cauchy horizon
may be destroyed, in contrast to the case illustrated in Figure
1.}
\end{figure}

In the following section, we define the class of space-times of
interest and obtain some useful relations for the metric functions
thereof. We consider spherically symmetric space-times admitting a
homothetic Killing vector field whose energy-momentum tensor obeys
the dominant energy condition. (A complete account of energy
conditions in spherical symmetry is given in the appendix.) For
generality, no further restrictions are imposed at this stage,
although some differentiability conditions at the past null cone
of the scaling origin and at the Cauchy horizon will be imposed.
Using co-ordinates adapted to the homothety and to the past null
cones of the central world-line, simple conditions can be given on
the metric which determine the visibility or otherwise of the
singularity at the scaling origin $\so$. This allows a simple way
of identifying both the past null cone $\pnc$ of $\so$ and the
Cauchy horizon $\ch$. In Section 3, we determine the behaviour of
completely general time-like geodesics (i) crossing $\pnc$ and
(ii) crossing the Cauchy horizon. These are used to calculate
fluxes of the scalar field at the respective surfaces. The
minimally coupled scalar wave equation is studied in the next
section. A mode decomposition relying on the Mellin transform is
used, and the asymptotic behaviour of the general solution at
$\pnc$ is determined. This is used to impose the boundary
condition that an arbitrary observer with unit time-like tangent
$v^a$ measures a finite flux
$\left.v^a\nabla_a\Phi\right|_{\pnc}$. We also demand that the
influx at ${\cal{J}}^-$ be finite. The modes not ruled out by
these boundary conditions are then allowed to evolve up to the
Cauchy horizon and the flux $\left.v^a\nabla_a\Phi\right|_{\ch}$
is calculated. Our principal result is that this flux is finite
for all the cases we consider.

\section{Self-similar spherically symmetric space-times admitting
a naked singularity.} We will consider the class of space-times
which have the following properties. Space-time $(\cm,g)$ is
spherically symmetric and admits a homothetic Killing vector
field. These symmetries pick out a scaling origin $\so$ on the
central world-line $r=0$ (which we will refer to as the axis),
where $r$ is the radius function of the space-time. We assume
regularity of the axis to the past of $\so$ and of the past null
cone $\pnc$ of $\so$. We will use advanced Bondi co-ordinates
$(v,r)$ where $v$ labels the past null cones of $r=0$ and is taken
to increase into the future. Translation freedom in $v$ allows us
to situate the scaling origin at $(v=0,r=0)$ and identifies $v=0$
with $\pnc$. The homothetic Killing field is
\[ {\vec{\xi}}=v\frac{\partial}{\partial v}+r\frac{\partial}{\partial
r}.\] The line element may be written \be ds^2 =
-2Fe^{2\psi}dv^2+2e^\psi dvdr+r^2d\Omega^2,\label{lel} \ee where
$d\Omega^2$ is the line element of the unit 2-sphere. The
homothetic symmetry implies that $F(v,r)=F(x), \psi(v,r)=\psi(x)$
where $x=v/r$. The only co-ordinate freedom remaining in
(\ref{lel}) is $v\to V(v)$; this is removed by taking $v$ to
measure proper time along the regular center $r=0$.

We will not specify the energy-momentum tensor of $(\cm,g)$, but
will demand that it satisfies the dominant energy condition. A
complete description of energy conditions in spherical symmetry is
given in Appendix A. Of these, the following will be used. (These
are equations (\ref{sec1}), (\ref{sec2}) and (\ref{dec6})
respectively.)
\begin{eqnarray} x\psi^\prime \leq 0,\label{ec1}
\\e^\psi(F^\prime+xF^2e^\psi\psi^\prime)\leq
0,\label{ec2}\\
1-2F+2x(F^\prime+F\psi^\prime)\geq 0.\label{ec3}
\end{eqnarray}

We impose the following regularity conditions at the axis. As
previously mentioned, we take $v$ to be proper time along the axis
for $v<0$. Noting that $x\to-\infty$ on this portion of the axis,
(\ref{lel}) then gives \be \lim_{x\to-\infty} 2Fe^{2\psi} =
1.\label{reg1}\ee The other regularity condition that we use is
that all curvature invariants are finite on $r=0,v<0$. In the
present case, the (invariant) Misner-Sharp mass is given by
\[ E=\frac{r}{2}(1-2F).\]
Then $E/r^3$ is a curvature invariant; this term has the same
units as e.g.\ $R$ and $\Psi_2$. Demanding that $E/r^3$ be finite
on the axis yields \be \lim_{x\to-\infty}
F=\frac12.\label{reg2}\ee Combining (\ref{reg1}) and (\ref{reg2})
gives these regularity conditions: \be F(-\infty)=\frac12,\quad
\psi(-\infty)=0.\label{reg}\ee

We define the {\em interior region} $\cm_{int}$ of space-time to
be the interior of $\pnc$, i.e.\ the interior of the causal past
of $\so$. The {\em exterior region} $\cm_{ext}$ is defined to by
$\cm=\cm_{int}\cup\pnc\cup\cm_{ext}$. (These definitions are in
line with those of \cite{christo1}.) We assume that the metric is
regular throughout $\cm_{int}\cup\pnc$ - this set does not include
$\so$ - by which we mean $F,\psi\in C^2(-\infty,0]$. As a Cauchy
horizon can only form in $\cm_{ext}$, we assume further that
$F,\psi\in C^2(-\infty, x_*)$ for some $x_*>0$. As we will see, if
a Cauchy horizon develops, it must be of the form $x=x_c$ for some
$x_c>0$. Our assumption is that the metric is regular at least up
until the Cauchy horizon.

Since we are studying collapse, our assumptions must include some
statement of regularity - in the sense of the absence of trapped
surfaces - of an initial configuration. The 2-sphere $(v,r)$ is
trapped if and only if
\[ \chi(v,r):=g^{ab}\nabla_ar\nabla_br <0.\]
In the present case, this is equivalent to $F<0$, and implies that
the condition for an apparent horizon is $F=0$. So in order to
express the notion that the matter is initially in some
non-extreme state, we rule out trapped or marginally trapped
surfaces in the interior region $\cm_{int}$. We will also demand
that $\pnc$ is not foliated by marginally trapped surfaces, and so
we take
\[ F(x)>0 \quad \hbox{for all}\quad x\in(-\infty,0].\]

Next, we point out the inevitability of there being a curvature
singularity at $\so$. Any curvature invariant which has units
$L^{-2}$ is of the form $c(x)r^{-2}$. For example,
\[ \frac{E}{r^3}=\frac{1-2F}{2r^2}.\]
This term diverges as we approach $\so$ along the null line $x=0$
unless $F(0)=\frac12$. But subject to the assumption that $F>0$
for $x<0$, we see that the surfaces $x=x_c<0$ are time-like. So we
may also approach $\so$ along $x=x_c<0$, and we then see that
$E/r^3$ diverges unless $F\equiv\frac12$ on $(-\infty,0]$.
Applying the same reasoning to the invariant
\[
\frac{E}{r^3}+\Psi_2+\frac{R}{12}=\frac{1}{2r^2}(1-2F+2x(F^\prime+F\Psi^\prime)),\]
regularity at $\so$ would require $\Psi\equiv 0$ on $(-\infty,0]$
(we have used the boundary condition (\ref{reg}) here). Hence
$\cm_{int}$ is a portion of flat space-time. So avoiding the
trivial case implies the existence of a curvature singularity at
$\so$.

Let us now prove the assertion above regarding the Cauchy horizon.

\begin{prop}
Let $x_c$ be the first positive root of $G(x):=Fe^\psi=\frac1x$,
if such exists and $x_c=+\infty$ otherwise. Then there are no
future pointing outgoing radial null curves emanating from $\so$
in the region $x\in(0,x_c)$.
\end{prop}

\noindent{\bf Proof} The outgoing radial null curves of
(\ref{lel}) satisfy \be
\frac{dr}{dv}=F(x)e^\psi(x)=G(x).\label{rngeq}\ee Let $(v_i,r_i)$
be a point on a solution curve $\gamma_i$ of (\ref{rngeq}) in the
region $0<x<x_c$. Then $x_i=v_i/r_i<x_c$, and so
$G(x_i)<\frac{1}{x_i}$. If $x_c$ is finite, we note that $x=x_c$
is a solution of (\ref{rngeq}),and so by uniqueness, $\gamma_i$
cannot cross $x=x_c$ away from $\so$ i.e.\ for $v>0$. Thus
\[ \left.\frac{dx}{dv}\right|_{\gamma_i}=\frac1r(1-xG)>0\]
for $v\in(0,v_i]$. Note that this inequality is immediate when
$x_c=+\infty$. So the inequality applies generally and says that
as $v\downarrow 0$, $x$ decreases and is bounded below by $0$.
Hence the limit
\[ x_l=\lim_{v\downarrow0}\left.x(v)\right|_{\gamma_i}<x_c\]
exists and is non-negative. Thus either $r\to r_*>0$ as
$v\downarrow0$ - in which case the singularity is avoided - or
$r\to 0$ in the limit. In this case,
\begin{eqnarray*}
x_l&=&\lim_{v\downarrow0} \frac{v}{r}\\
&=&\lim_{v\downarrow0}\frac{1}{r^\prime(v)}\\
&=&\lim_{v\downarrow0}\frac{1}{G(x)}=\frac{1}{G(x_l)},\end{eqnarray*}
where all limits are taken along $\gamma_i$ and l'Hopital's rule
is used in the second line. The conclusion that $x_l<x_c$ is a
root of $xG=1$ contradicts minimality of $x_c$ and completes the
proof.\hfill$\Box$

\begin{corr}
If $G(x)<\frac1x$ for all $x>0$, then the singularity $\so$ is
censored. \hfill$\Box$\end{corr}

\begin{corr} If $G(x)=\frac1x$ for some values of $x>0$, then
$x=x_c$ is the Cauchy horizon $\ch$ of the space-time, where $x_c$
is the smallest positive root of $xG=1$.\hfill$\Box$
\end{corr}

These results show an advantage of describing self-similar
collapse in the co-ordinates $v$ and $r$: the visibility of the
singularity at $\so$ (and indeed the presence of an apparent
horizon $F=0$) can be read-off from the metric. More accurately,
the presence of a naked singularity can be determined by tracking
the evolution of metric functions, and without having to integrate
geodesic equations.

An apparent horizon may form either before or after the Cauchy
horizon. This horizon must be space-like, and the region lying to
its future is trapped:

\begin{prop}
If $F(x_a)=0$ for some $x_a>0$, then $x=x_a$ is space-like and the
region $x\geq x_a$ is trapped or marginally trapped.
\end{prop}
\noindent{\bf Proof} Restricting to $\Sigma_a:\{v=x_a r\}$ in
(\ref{lel}) gives
\[ \left.ds^2\right|_{\Sigma_a}=
2x_ae^{\psi(x_a)}(1-x_aG(x_a))dr^2+r^2d\Omega^2,\] which has
spatial signature at $G=F=0$ when $x_a>0$. From (\ref{ec2}), we
see that $F^\prime\leq 0$ at an apparent horizon. Hence $F(x)\leq
0$ for $x\geq x_a$.\hfill$\Box$

We conclude this section with a lemma which will play a central
role in determining the stability of $\ch$ with respect to scalar
radiation.

\begin{lemma}
$G^\prime< 0$ prior to the formation of a Cauchy horizon.
\end{lemma}
\noindent{\bf Proof} We note first that the results of
Propositions 1 and 2 show that $0<xG<1$ for $x\in(0,x_c)$. Then
(\ref{ec1}) gives
\[ xGF\psi^\prime=xF^2e^\psi\psi^\prime>F\psi^\prime,\]
and using (\ref{ec2}) we get
\[
F^\prime+F\psi^\prime<F^\prime+xF^2e^\psi\psi^\prime \leq 0,\]
i.e.\ $G^\prime(x)< 0$ for $x\in(0,x_c)$.\hfill$\Box$

\begin{corr}
$G^\prime(x_c)\leq 0$.\hfill$\Box$
\end{corr}

We note that if $G^\prime(x_c)=0$, then
$\left.R_{ab}k^ak^b\right|_{\ch}=0$, where $k^a$ is tangent to the
outgoing radial null direction. This implies that there is no
ingoing radiative flux of energy-momentum crossing the Cauchy
horizon. We rule out this situation as being physically
unrealistic and so we will assume that $G^\prime(x_c)<0$.

\section{Time-like geodesics crossing $\pnc$ and $\ch$.}
The stability of the Cauchy horizon will be studied from the point
of view of the behaviour of the flux of scalar radiation measured
by an observer crossing the horizon. This flux is
$\cf=v^a\nabla_a\Phi$, where $\Phi$ is the scalar field and $v^a$
is the unit tangent to an arbitrary time-like geodesic. Thus we
will need to determine the behaviour of the tangent $v^a$ for such
arbitrary geodesics at the Cauchy horizon. Since we will impose
boundary conditions on $\Phi$ in terms of the fluxes at $\pnc$, we
will need to do the same at this surface. The full set of
equations governing time-like geodesics may be written in the form
\begin{eqnarray}
{\ddot v}-\frac1r(x(G^\prime+G\psi^\prime)-\psi^\prime){\dot
v}^2-\frac{e^{-\psi}}{r^3}L^2,\label{tlg1}&=&0\\
-2Ge^\psi{\dot v}^2+2e^\psi{\dot v}{\dot
r}+\frac{L^2}{r^2}&=&-1,\label{tlg2}\\
{\dot{\Omega}}&=&\frac{L}{r^2},\label{tlg3}
\end{eqnarray}
where the overdot represents differentiation with respect to
proper time $\tau$, $L$ is the conserved angular momentum and
$\Omega$ is an azimuthal angular variable. (\ref{tlg3}) plays no
further role below, but is given for completeness. It is
convenient to rewrite (\ref{tlg1}) and (\ref{tlg2}) as a first
order system. Defining $X=(r,v,u)^T$ where $u:={\dot v}$, these
equations may be written as \be {\dot X}=H(X)=
\left(\begin{array}{c}
  \frac{e^{-\psi}}{2u}(2Ge^\psi u^2-\frac{L^2}{r^2}-1) \\
  u \\
  \frac1r(x(G^\prime+G\psi^\prime)-\psi^\prime)u^2+\frac{L^2}{r^3}e^{-\psi}
\end{array}
\right). \label{fos} \ee

A future-pointing time-like geodesic crossing $\pnc$ corresponds
to a solution of (\ref{fos}) with initial values $r_0>0$, $v_0=0$,
$u_0>0$. The assumptions of the previous section indicate that $H$
is $C^1$ in a neighbourhood of $(r_0,v_0,u_0)\in \mathbb{R}^3$,
and so standard theorems imply the existence of a $C^1$ solution
for $X$ which exists for (at least) finite duration. Note that
this implies that both $v$ and $r$ (via (\ref{tlg2})) are $C^2$
functions of proper time $\tau$ in a neighbourhood of $\pnc$. Thus
we can apply Taylor's theorem and write
\cite{def1}\begin{eqnarray*}
v(\tau)&=&u_o\tau+v_2\tau^2+O(\tau^3),\\
r(\tau)&=&r_0+r_1\tau+O(\tau^2),
\end{eqnarray*}
where the coefficients $v_2,r_1$ can be given in terms of the
initial data and metric functions and we have set $\tau=0$ at
$\pnc$. From this we may write down the following result which
will be required below. \begin{prop} For any future-pointing
time-like geodesic crossing $\pnc$, we have
\begin{eqnarray} v&\sim &u_0\tau,\quad
{\dot v}\sim u_0,\label{nsol1}\\
x&\sim &\frac{u_0}{r_0}\tau,\quad {\dot
x}\sim\frac{u_0}{r_0},\label{nsol2} \end{eqnarray} as $\tau\to 0$
where $\tau=0$ on the geodesic at $\pnc$.\hfill$\Box$\end{prop}

Obtaining equivalent results at the Cauchy horizon is more
difficult, as this corresponds to a singular point of the geodesic
equations. Two things must be established: the existence of
time-like geodesics crossing the horizon and the limiting values
of the components of the tangent vector at the horizon. The proof
below requires an assumption on the level of differentiability at
the horizon which it would be desirable to remove.

\begin{prop} Suppose that $G$ and $\psi$ are differentiable at
$x=x_c$. Then all radial time-like geodesics whose initial points
are sufficiently close to the Cauchy horizon will cross the
horizon in finite time. For any time-like geodesic crossing the
horizon, the components of the tangent $\dot{x}$ and $\dot{v}$
have finite non-zero values at the horizon which, denoting them by
$\dot{x}_c$ and $\dot{v}_c$ respectively, satisfy the relation \be
\dot{v}_c=\frac12\frac{x_c^2}{v_c\dot{x}_c}e^{-\psi_c},\label{glims}\ee
where the subscript refers to the value of a quantity at $x=x_c$.
\end{prop}

{\bf Proof} (i) First, we establish a first order non-autonomous
system for the geodesics. If $\xi^a$ is the homothetic Killing
vector field and $u^a$ is tangent to a time-like geodesic, then
\[ \frac{d}{d\tau}(\xi_au^a)=-1,\]
where $\tau$ is proper time along the geodesic (see e.g.\ Appendix
C of \cite{wald}). Integrating yields
\[
g_{ab}\xi^au^b=2\frac{v}{x}e^\psi(1-xG){\dot{v}}-\frac{v^2}{x^2}e^\psi\dot{x}=k-\tau,\]
for some $k$ which is constant along the geodesic. Combining with
\be \frac{2}{x}e^\psi(1-xG){\dot{v}}^2-\frac{2}{x^2}e^\psi
v\dot{v}\dot{x}=-(1+\frac{L^2}{r^2})\label{tlg2a}\ee (which is
(\ref{tlg2}) written in terms of $v$ and $x$) we obtain the first
order system
\begin{eqnarray} {\dot{x}}&=&\pm
\frac{x^2}{v^2}e^{-\psi}Y^{1/2},\label{fog1}\\
{\dot{v}}&=&\frac12\frac{x}{v}e^{-\psi}(1-xG)^{-1}(-(\tau-k)\pm
Y^{1/2}),\label{fog2}\end{eqnarray} where
\[ Y=(\tau-k)^2+2\frac{v^2}{x}(1-xG)e^\psi(1+\frac{L^2}{r^2}).\]
We choose the upper sign, which corresponds to future-pointing
geodesics.

(ii) For radial $(L=0)$ time-like geodesics, (\ref{tlg1}) becomes
\[
{\ddot{v}}=(x^2G^\prime+x\psi^\prime(xG-1))\frac{{\dot{v}}^2}{v}.\]
Since $G^\prime(x_c)<0$, the coefficient on the right hand side is
negative for values of $x<x_c$ sufficiently close to $x_c$. Hence
a geodesic with initial value $x_0=x(\tau_0)$ sufficiently close
to $x_c$ in this sense satisfies ${\ddot{v}}<0$ for
$\tau\geq\tau_0$, and so $v$ cannot diverge to infinity in finite
time.

(iii) Next, we establish that if $v\to\infty$ as $\tau\to\infty$
along a geodesic which does not cross the Cauchy horizon, then
$x\to x_c$ as $\tau\to\infty$. From (\ref{tlg2a}), we can write
\[ \frac{\dot{v}}{v}<\frac{\dot{x}}{x(1-xG)}.\]
Integrating both sides yields and taking $v=v_0$ at some
$0<x_0<x_c$, we get
\[ v<v_0\exp\left(\int_{x_0}^x \frac{dy}{y(1-yG)}\right)\]
for $x_0<x$. Thus if $v$ diverges to $+\infty$, then so too must
the integral. This can only occur if the integrand diverges, i.e.\
if $x\to x_c$. Now we show that provided a geodesic has initial
point sufficiently close to $x=x_c$, it cannot behave in this way.

(iv) Consider a radial time-like geodesic for which $v\to\infty$
and $x\to x_c$ as $\tau\to\infty$. We have from (\ref{tlg1})
\[ \frac{v\ddot{v}}{\dot{v}^2}\to x_c^2G^\prime(x_c)=-k^2<0\]
as $\tau\to\infty$. Integrating and reusing this relation yields
the asymptotic relations
\begin{eqnarray}
\dot{v}&\sim & c_1v^{-k^2},\label{asm1}\\
\ddot{v}&\sim & - k^2c_1v^{-2k^2-1}\label{asm2}
\end{eqnarray}
as $\tau\to\infty$ for some $c_1>0$. Using these and (\ref{tlg2a})
we obtain \begin{eqnarray} \dot{x}& \sim & c_2
v^{k^2-1}\label{asm3}\end{eqnarray} as $\tau\to\infty$ for some
$c_2>0$. We must have $\dot{x}\to0$ as $\tau\to\infty$, for
otherwise $\dot{x}$ is positive and bounded away from zero for an
infinite amount of time and so $x$ reaches $x_c$ in finite time.
Our present assumption is that this does not happen, so we must
have $k^2<1$.

The geodesic equations yield \be
2A\ddot{v}+vB\ddot{x}+2A^\prime\dot{x}\dot{v}+vB^\prime\dot{x}^2=0,\label{tlg4}\ee
where
\[ A(x)=\frac2xe^\psi(1-xG),\quad B=-\frac{2}{x^2}e^\psi.\]
Using the assumption that these terms are differentiable at the
Cauchy horizon, we have from (\ref{asm1})-(\ref{asm3}),
\cite{def1}
\begin{eqnarray*}
2A\ddot{v}&\sim&-2Ak^2c_1^2v^{-2k^2-1}=o(v^{-2k^2-1}),\\
2A^\prime\dot{x}\dot{v}&\sim&2A^\prime c_1c_2v^{-1}=O(v^{-1}),\\
vB^\prime\dot{x}^2&\sim& B^\prime v^{2k^2-1}.
\end{eqnarray*}
Comparing these with (\ref{tlg4}), we see that we must have
\[ \lim_{v\to\infty}
v\ddot{x}=-\lim_{v\to\infty}v\frac{B^\prime}{B}\dot{x}^2.\] We
have
\[ \frac{B^\prime}{B}=-2x^{-1}+\psi^\prime.\]
Using the energy condition (\ref{ec1}), we see that this term is
strictly negative at the Cauchy horizon, and so this implies that
$\ddot{x}$ is positive for sufficiently large values of $\tau$.
However this contradicts the fact that $\dot{x}(\tau)>0$ with
$\dot{x}\to 0$ as $\tau\to 0$. Hence the geodesic cannot extend to
arbitrarily large values of $v$ without first crossing the Cauchy
horizon.

(v) To conclude, it has been established that (at least some)
radial geodesics cross the horizon in finite time and so with a
finite value $v_c$ of $v$. For any such geodesic, including
non-radials,  we can read off from (\ref{fog1}) the non-zero and
finite limiting value of $\dot{x}$ and from (\ref{tlg2a}) we
obtain (\ref{glims}).\hfill${\Box}$

\section{The scalar field on the Cauchy horizon.}
Now we are in a position to examine the stability of the Cauchy
horizon by measuring the flux of the scalar field in different
regions of the space-time.\\In order to measure the flux
$\mathcal{F}=v^a \nabla_a \Phi$ we need first the solution of the
scalar wave equation,
\[
\Box\Phi=(-g)^{-\frac{1}{2}} \partial_a \left[ (-g)^{\frac{1}{2}}
g^{ab} \partial_b \Phi \right]=0.
\]
We exploit the spherical symmetry of the space time and split the
scalar field,
\[
\Phi(v,x,\theta,\phi)=T(v,x)A(\theta,\phi),
\]
where we use the advanced null coordinate $v$, the homothetic
coordinate $x$, and the standard angular coordinates
$\theta,\phi$. Then the line element in these co-ordinates reads
\[
ds^2=2e^{\psi} \left( \frac{1}{x}-G \right) dv^2-\frac{2e^{\psi}
v}{x^2} dvdx + \frac{v^2}{x^2}d\Omega^2.
\]
By using separation of variables we arrive at a p.d.e. in $v,x$
\begin{eqnarray}
2x^2 \left( \frac{1}{x}-G \right) T,_{xx}+2v T,_{xv}- 2x^2 G' T,_x
\nonumber \\ - 2vT,_v - \rho e^{\psi} T=0 \label{vn}
\end{eqnarray}
where $\rho=l(l+1)$ is the separation constant, $l=0,1,2\ldots$ is
the multipole mode number, and $'$ denotes differentiation
w.r.t.$\:x$. The complementary p.d.e. in $\theta,\phi$ reduces to
a form of Legendre's equation and is solved by the spherical
harmonic
functions, $P^m_l(\theta,\phi)$.\\
We can perform a Mellin transformation on the field, defined by
\[
M\{T\}(x,n)=H_n(x)=\int_0^{\infty} T(v,x) v^{n-1} dv
\]
which amounts to replacing $T(v,x)$ with $v^n H_n(x)$, where $n$
is an as yet unconstrained complex parameter. Equation (\ref{vn})
thus reduces to an o.d.e. in $H(x)$,
\begin{eqnarray}
2x^2 \left( \frac{1}{x}-G \right) H'' + ( 2n- 2x^2G') H' \nonumber
\\ - \left( \frac{2n}{x}+ \rho e^{\psi} \right) H =0 \label{od}
\end{eqnarray}
where we have suppressed the subscript $n$. Performing the inverse
Mellin transform on the solution of this o.d.e. over a
contour in the viable range of $n$ will return the solution to (\ref{vn}).\\
This o.d.e. has a number of singular points, namely $x=0$ and the
roots of $xG=1$, the lowest of which we have defined to be $x_c$.
The canonical form of a second order linear o.d.e. in a
neighborhood of $x=x_0$ is
\[
(x-x_0)^2 H''+(x-x_0) q(x)H'+p(x)H=0,
\]
and when we write equation (\ref{od}) in its canonical form in the
neighborhood of $x=0$, we find
\begin{eqnarray*}
q(x)=\frac{n-x^2 G'}{1-xG}, \qquad p(x)=-\frac{2n+\rho e^{\psi}
x}{2(1-xG)}.
\end{eqnarray*}
Since $q(x)$ and $p(x)$ are both $C^1$ in a neighborhood of $x=0$
we can use the method of Frobenius to solve (\ref{od}) on
$\mathcal{N}$ \cite{foot} (see e.g.\ Chapter 3 of \cite{BendOr}).
The indicial exponents are $1,-n$.\\As it stands we cannot make
any assumptions about $n$, however later analysis shows if
$-\textrm{Re}(n) \geq 1$ the flux of the scalar field will be
always infinite on $\mathcal{N}$, thus
we only consider $-\textrm{Re}(n)<1$.\\
It is possible for $1$ and $-n$ to differ by an integer and so the
method of Frobenius yields the following expression for the
solution to (\ref{od}) in a neighborhood of $x=0$,
\begin{eqnarray}
H(x)=c_1\sum_{m=0}^{\infty} a_m x^{m+1} \nonumber \\ +c_2\left\{k
\ln x \sum_{m=0}^{\infty}a_m x^{m_+1} + \sum_{m=0}^{\infty} b_m
x^{m-n} \right\}. \label{tf}
\end{eqnarray}
In this expression, $c_1$ and $c_2$ are arbitrary constants,
$a_0=b_0=1$ with $k=0$ if $1$ and $-n$ do not differ by an
integer, $a_0=1, b_0=0$ with $k=1$ if $1$ and $-n$ are equal, and
$a_0=b_0=1$ with $k\neq 0$ if $1+n=m$ for some positive integer
$m$.\\After some rearranging and some cancellations, the
expression for the flux on $\mathcal{N}$ is
\begin{eqnarray}
\mathcal{F}_1(v,r)=\dot{v} \sum_{m=0}^{\infty} a_m (m+n+1)
\frac{v^{n+m}}{r^{m+1}} \nonumber \\ - \dot{r} \sum_{m=0}^{\infty}
a_m (m+1)
\frac{v^{m+n+1}}{r^{m+2}} \\
\mathcal{F}_2(v,r)=\dot{v}\sum_{m=0}^{\infty} b_{m+1} (m+1)
\frac{v^{m}}{r^{m-n+1}} \nonumber \\ - \dot{r} \sum_{m=0}^{\infty}
b_m (m-n)
\frac{v^{m}}{r^{m-n+1}} \nonumber \\
-\dot{r}k \sum_{m=0}^{\infty} \left[ 1+ (m+1) \ln \left(
\frac{v}{r} \right) \right] a_m \frac{ v^{n+m+1}}{ r^{m+2}}
\nonumber \\+ \dot{v}k \sum_{m=0}^{\infty} \left[ 1+ (m+n+1) \ln
\left( \frac{v}{r} \right) \right] a_m \frac{ v^{m+n}}{ r^{m+1}}.
\end{eqnarray}
where the $1$ subscript denotes the $c_1$ part, and likewise the
$2$ subscript.\\
The components of the velocity have been shown to be finite on
$\mathcal{N}$ in Proposition 3, and we see that for the flux to
have a finite measure on $\mathcal{N}$, that is when $v=0$, we
require
\[
\textrm{Re}(n)> 0.
\]
Under this condition we let the scalar field evolve towards
$\mathcal{H}$, and examine its flux there.\\ \underline{Note}: A
scalar field coming from past null infinity will have a finite
flux thereon if $\textrm{Re}(n) \leq 1$. While this physically
desirable condition should be imposed, it
does not play any role in later analysis.\\
\\
When we write (\ref{od}) in its canonical form around $x=x_c$, we
find the coefficients are
\begin{eqnarray*}
q(x)=\left(\frac{x-x_c}{x}\right)\left(\frac{n-x^2 G'}{1-x
G}\right) \\ p(x)=\left(\frac{x-x_c}{x}\right)^2
\left(\frac{n+\frac{\rho e^{\psi} x}{2}}{1-xG}\right).
\end{eqnarray*}
Now we reach an important distinction, whether $G(x)$ has a
$\textit{unique}$ lowest root or $\textit{multiple}$ lowest roots.
We distinguish the two cases so:
\begin{lemma} When $xG=1$ has a
unique lowest root,
\begin{eqnarray*}
x_c^2G'(x_c)+1>0.
\end{eqnarray*}
When $xG=1$ has a multiple lowest root, \begin{eqnarray*}
x_c^2G'(x_c)+1=0.
\end{eqnarray*}\hfill$\Box$
\end{lemma}
The two cases will lead to very different analyses, thus we treat
them separately.\\
\\
(i) The first case leads to $q(x),p(x)$ being $C^1$ on $x=x_c$,
thus $x_c$ is a regular singular point and hence we can use the
method of Frobenius. The indicial exponents are $0,1-q_0$ where
\begin{eqnarray*}
q_0=\frac{x_c^2 G'(x_c)-n}{x_c^2 G'(x_c)+1}.
\end{eqnarray*}
Since $n>0$, Lemma $1$ and $2$ tell us $q_0 <0$, hence $1-q_0
>0$, which gives us
\begin{eqnarray}
H(x)=C_1\sum_{m=0}^{\infty} A_m \zeta ^ {m+1-q_0} \nonumber \\
+C_2\left\{k \ln \zeta \sum_{m=0}^{\infty}A_m \zeta ^ {m+1-q_0} +
\sum_{m=0}^{\infty} B_m \zeta^m \right\}
\end{eqnarray}
where $\zeta=x-x_c$, and the coefficients have the same structure
as (\ref{tf}). From this we calculate the flux,\\ \vspace{-0.1in}
\begin{eqnarray}
\mathcal{F}_1=\dot{x}v^n C_1 \sum_{m=0}^{\infty}(m+1-q_0) A_m
\zeta^{m-q_0} \nonumber \\+ \dot{v}nv^{n-1} C_1
\sum_{m=0}^{\infty} A_m \zeta^{m+1-q_0} \\
\mathcal{F}_2=\dot{x}v^n C_2 \Bigg[ k \sum_{m=0}^{\infty} A_m
\left[ \ln \zeta(m+1-q_0) +1 \right]
\zeta^{m-q_0} \nonumber \\ +\sum_{m=0}^{\infty} B_m m \zeta^{m-1} \Bigg] \nonumber\\
+ \dot{v}nv^{n-1} C_2 \left[ \sum_{m=0}^{\infty} B_m \zeta^m + k
\ln \zeta \sum_{m=0}^{\infty} A_m \zeta^{m+1-q_0} \right].
\end{eqnarray}
Using the finiteness of $\dot{v},\dot{x}$ given in Proposition 4,
we see that if $q_0 <0$, that is if $n>0$, this expression is
finite on $\mathcal{H}$, i.e. when $x-x_c=\zeta=0$.
\\Thus in the case of $xG=1$ having a unique lowest root, a scalar
field measuring a finite flux entering the region will measure a
finite flux on the
Cauchy horizon.\\
\\
(ii) If $x_c^2 G'(x_c)+1=0$, $x_c$ is an irregular singular point
of (\ref{od}). Note that this is a special case which one would
expect to correspond to a set of measure zero in the class of
space-times under consideration. We label $\eta=x_c-x$ and examine
solutions to the o.d.e.\ in the asymptotic limit $\eta \downarrow
0$ (see e.g. Chapter 3 of \cite{BendOr}).\\We assume the solution
to (\ref{od}) can be written in the form
\[
H(\eta)= e^{S(\eta)},
\]
reducing (\ref{od}) to an o.d.e. in $S$. Now we assume the common
property near irregular singular points,
\[
\ddot{S}=o(\dot{S}^2), \qquad \edo
\]
where the overdot denotes differentiation with respect to $\eta$.
(\ref{od}) becomes a quadratic in $\dot{S}$,
\begin{eqnarray}
\dot{S}^2 \left\{ (x_c-\eta)-(x_c-\eta)^2 G
\right\}-\left(n+(x_c-\eta)^2 \dot{G} \right) \dot{S} \nonumber \\
\sim \frac{n}{x_c-\eta}+\frac{\rho e^{\psi}}{2}, \qquad \edo
\end{eqnarray}
If we consider $xG=1$ to have a lowest root of multiplicity $k$,
then we can write its Taylor series around $\eta=0$ as
\[
1-(x_c-\eta)G(\eta)=P(\eta)=\eta^k
\frac{P^{(k)}(0)}{k!}+O(\eta^{k+1})
\]
This means if the lowest root is of multiplicity $k$, we need the
metric functions to be $C^k$. This is not too much of a
restriction however, since the class of functions with roots of
multiplicity $k$ becomes very small as $k$ increases, meaning we
are dealing with a very special case in this analysis.\\ We can
make the approximation {\setlength\arraycolsep{2pt}
\begin{eqnarray*}n+(x_c-\eta)^2
\dot{G} &\sim& n+1, \qquad \edo,
\end{eqnarray*}}
and since we assume the metric coefficients are at least $C^2$, we
can approximate $e^{\psi}$ by the first term in its expansion,
$c_0$, in the limit $\edo$. Thus we arrive at a quadratic in $S$,
\begin{eqnarray}
\eta^k(\dot{S})^2-\alpha\dot{S} \sim \beta, \qquad \edo, \qquad\qquad\\
\nonumber \\
\alpha=\frac{k!(n+1)}{x_cP^{(k)}(0)},\quad
\beta=\frac{k!}{x_cP^{(k)}(0)}\left( \frac{n}{x_c}+\frac{\rho
c_0}{2} \right) \nonumber
\end{eqnarray}
where $\alpha,\beta>0$ (if $Re(n)>0$) and constant in the limit
$\edo$, and $k>1$. This quadratic has two solutions corresponding
to two linearly independent solutions of (\ref{od}), which are
\begin{eqnarray*}
S_1&\sim&-\frac{\alpha}{(k-1)}\eta^{1-k}+O(\eta)
\\S_2&\sim&-\frac{\beta}{\alpha}\eta+\frac{\beta^2}{\alpha^3}\frac{\eta^{k+1}}{(k+1)}
+O(\eta^{2k+1}), \qquad \edo.
\end{eqnarray*}
At this point we verify our earlier assumption, namely
\begin{eqnarray*}
\ddot{S}=o(\dot{S}^2), \qquad \edo.
\end{eqnarray*} Thus we have constructed two solutions to (\ref{od}),
\begin{eqnarray}
H_1(\eta)&=&\eta^k\exp\left\{-\frac{\alpha}{(k-1)}\eta^{1-k}+O(\eta)\right\}\\
H_2(\eta)&=&\exp\{O(\eta)\}
\end{eqnarray}
Both of these functions and their derivatives are finite in the
limit $\edo$, $x \rightarrow x_c$ if $Re(n)>0$, and thus the
resulting expressions for the flux are finite, where again we use
Proposition 4.\\We summarize thus:

\begin{prop}
Let space-time ($\mathcal{M}$,g) satisfy the requirements of
$\emph{Section II}$ and admit a Cauchy horizon $x=x_c$. Assume
also that $g_{ab} \in C^2$ at $x=x_c$. Then a scalar field which
has a finite flux on $\mathcal{N}$, the past null cone of
$\mathcal{O}$, will also have a finite flux on the Cauchy horizon,
$\mathcal{H}$.
\end{prop}

\section{Conclusions}
We have shown that the Cauchy horizon formed by collapse in a self
similar, spherically symmetric space-time is stable with respect
to scalar radiation. This space-time is very general, the only
other constraints being that the field satisfies the dominant
energy condition, and, other than the special case discussed in
Section IV(ii), we require the metric functions to be $C^2$ on
$\pnc$ and $\ch$. These differentiability conditions are stronger
than one would like to assume ({\em cf.}\ the $C^0$ Cauchy
horizons appearing in the collapse of wave maps \cite{bizon}), but
are as low as one can go without having to resort to a generalised
solution concept for the wave equation.
\\The next step is to examine whether linear perturbations of the
metric functions will lead to an unstable Cauchy horizon, as is
seen, for example, in the Reissner-Nordstr\"om solution. Such an
examination would be more significant in considering cosmic
censorship. Is it difficult to anticipate the general outcome of
such an examination. One expects to observe instability for the
case of a massless scalar field \cite{christo2}, but stability for
(some sectors of) perfect fluid collapse \cite{harada}. The
present results and the Cauchy horizon stability conjecture would
lead one to expect stability in general \cite{debbie}.

\section*{Acknowledgement}
This research is supported by Enterprise Ireland grant
SC/2001/199.

\appendix
\section{Energy conditions in spherical symmetry}
\subsection{Spherical symmetry} We write the line element in double
null coordinates;
\[
ds^2=-2e^{-2f}dudv+r^2d\Omega^2,\] where $f=f(u,v)$, $r=r(u,v)$
and $d\Omega^2$ is the line element on the unit 2-sphere. The
non-vanishing Ricci tensor terms are
\begin{eqnarray*}
R_{uu}&=&-2r^{-1}(r_{uu}+2r_uf_u),\\
R_{vv}&=&-2r^{-1}(r_{vv}+2r_vf_v),\\
R_{uv}&=&-2r^{-1}(r_{uv}-rf_{uv}),\\
R_{\theta\theta}&=&\csc^2\theta
R_{\phi\phi}=2\frac{E}{r}+2e^{2f}rr_{uv},
\end{eqnarray*}
where $E$ is the Misner-Sharp mass,
\[ E=\frac{r}{2}(1+2e^{2f}r_ur_v).\]
Subscripts on $f,r$ denote partial derivatives. The only
non-vanishing Weyl tensor term is
\begin{eqnarray*}
\Psi_2&=&-\frac{1}{3}\frac{E}{r^3}+\frac13e^{2f}(f_{uv}+r^{-1}r_{uv})\\
&=&-\frac{E}{r^3}-\frac{1}{12}g^{AB}R_{AB}+\frac{R_{\theta\theta}}{3r^2},
\end{eqnarray*}
where $x^A$ are co-ordinates in the Lorentzian 2-space.

\subsection{The strong energy condition}

Our aim is to write down a set of conditions on the curvature
terms listed above which are equivalent to the strong energy
condition:
\[ R_{ab}v^av^b\geq 0\]
for all (future-pointing) causal vectors ${\vec v}$.

\subsubsection{Null vectors} The radial null directions are
$\delta^a_u, \delta^a_v$. These give
\[
R_{uu}\geq 0,\qquad R_{vv}\geq 0.
\]

At any point, the general non-radial null direction may be written
as \begin{equation} v^a =
\alpha\delta^a_u+\beta\delta^a_v+\gamma\delta^a_\phi.\label{vec}\end{equation}
The null condition is
\[ \alpha\beta = 2r^2e^{2f}\gamma^2.\]
We find
\[ R_{ab}v^av^b =
\alpha^2R_{uu}+\beta^2R_{vv}+2\alpha\beta(2f_{uv}+2\frac{E}{r^3}e^{-2f}).\]
This is non-negative for all non-radial null vectors if and only
if it is non-negative for all values of $\alpha, \beta$ with
$\alpha\beta>0$. In turn, this is true if
\[ \min_{\mu>0} Q(\mu) \geq 0,\]
where $\mu=\alpha/\beta > 0$ and
\[
Q(\mu)=\mu^2R_{uu}+2\mu(2f_{uv}+2\frac{E}{r^3}e^{-2f})+R_{vv}.\]

If $R_{uu}=R_{vv}=0$, this is simply
\[ f_{uv}+\frac{E}{r^3}e^{-2f}\geq 0.\]
If $R_{uu}=0$ and $R_{vv}\neq 0$, the condition is equivalent to
\[ 2\mu(2f_{uv}+2\frac{E}{r^3}e^{-2f})+R_{vv}\geq0\]
for all $\mu>0$. This is satisfied if $
f_{uv}+\frac{E}{r^3}e^{-2f}\geq 0$. If $
f_{uv}+\frac{E}{r^3}e^{-2f}<0$, then the condition will be
violated for sufficiently large values of $\mu$ (which can always
be chosen). The same holds for $R_{uu}\neq0, R_{vv}=0$. Thus if
$R_{uu}R_{vv}=0$, the strong energy condition holds for null
directions if and only if \[ f_{uv}+\frac{E}{r^3}e^{-2f}\geq 0.\]

So now assume that $R_{uu}>0$, $R_{vv}>0$. The quadratic $Q(\mu)$
has a global minimum at
\[\mu_* = -2R_{uu}^{-1}(f_{uv}+\frac{E}{r^3}e^{-2f}),\]
while $Q(0)=R_{vv}>0$. Thus $Q(\mu)>0$ for $\mu>0$ if and only if
either $\mu_*\leq 0$ or $\mu_*>0$ and $Q(\mu_*)\geq 0$.

$\mu_*\leq 0$ if and only if $ f_{uv}+\frac{E}{r^3}e^{-2f}\geq 0$.

$\mu_*>0$ if and only if $f_{uv}+\frac{E}{r^3}e^{-2f}< 0$. In this
case,
\begin{eqnarray*}
Q(\mu_*)&=&-4R_{uu}^{-1}(f_{uv}+\frac{E}{r^3}e^{-2f})^2+R_{vv}
\geq 0\\
&\Leftrightarrow&|f_{uv}+\frac{E}{r^3}e^{-2f}|\leq\frac12(R_{uu}R_{vv})^{1/2}\\
&\Leftrightarrow&f_{uv}+\frac{E}{r^3}e^{-2f}\geq-\frac12(R_{uu}R_{vv})^{1/2}.
\end{eqnarray*}

Combining these results, we can say:

$R_{ab}v^av^b\geq 0$ for all null $v^a$ if and only if
\begin{eqnarray}
R_{uu}&\geq&0 \label{enc1}\\
R_{vv}&\geq&0 \label{enc2}\\
\frac12(R_{uu}R_{vv})^{1/2}+f_{uv}+\frac{E}{r^3}e^{-2f}&\geq&0.\label{enc3}
\end{eqnarray}

\subsubsection{Time-like vectors} Again we write
\[ v^a =
\alpha\delta^a_u+\beta\delta^a_v+\gamma\delta^a_\phi,\] and we can
use the time-like condition $g_{ab}v^av^b=-1$, so that
\[ \gamma^2=r^{-2}(2e^{-2f}\alpha\beta-1)\geq 0.\]
So in this case we are minimizing over the set
$\alpha\beta\geq\frac12 e^{2f}$. we do this by minimizing over the
hyperbola $\alpha\beta=c$ and then minimizing over all hyperbolas
$c\geq \frac12 e^{2f}$. This yields the conditions above and the
extra condition \begin{equation}
\frac12(R_{uu}R_{vv})^{1/2}+f_{uv}-r^{-1}r_{uv}\geq0.\label{enc4}
\end{equation}

\subsection{The weak energy condition} The weak energy condition
$T_{ab}v^av^b\geq 0$ for all causal $v^a$ can be written, using
Einstein's equation, as $R_{ab}v^av^b\geq \epsilon R/2$, where
$\epsilon=g_{ab}v^av^b$. ($R=$ Ricci scalar.) Thus the only extra
work to do is for time-like vectors. The algebra involved in the
previous section only needs minute changes, and we can show that
the weak energy condition is equivalent to (\ref{enc1}) -
(\ref{enc3}) and
\begin{equation}
\frac12(R_{uu}R_{vv})^{1/2}+r^{-1}r_{uv}+\frac{E}{r^3}e^{-2f}\geq0.\label{enc5}
\end{equation}

\subsection{The dominant energy condition} This states that for
every future-pointing timelike $v^a$, the vector $-T^{ab}v_b$ is
non-spacelike and future-pointing, and $T_{ab}v^av^b\geq 0 $.
Using the usual general form for $v^a$, we again have quadratic
inequalities for the parameters $\alpha$ and $\beta$ which can be
treated in the usual way. (The non-spacelike condition is
$g_{ac}T^{ab}T^{cd}v_bv_d\leq 0$; the left hand side is
homogeneous of degree 2 in $(\alpha,\beta)$ and so quadratic in
$\mu$.) The future-pointing condition is simple to examine by
assuming that $u,v$ increase into the future. The resulting
inequalities are
\begin{eqnarray*}
R_{uu}&\geq&0,\\
R_{vv}&\geq&0,\\
\frac{E}{r^3}e^{-2f}+r^{-1}r_{uv}&\geq&0,\\
\frac12(R_{uu}R_{vv})^{1/2}+\frac{E}{r^3}e^{-2f}+r^{-1}r_{uv}&\geq&|f_{uv}-r^{-1}r_{uv}|.
\end{eqnarray*}
Using the first three of these, we see that the left hand side of
the fourth is non-negative, and so the fourth is equivalent to the
{\em two} inequalities
\begin{eqnarray*}
\frac12(R_{uu}R_{vv})^{1/2}+\frac{E}{r^3}e^{-2f}+2r^{-1}r_{uv}-f_{uv}&\geq&0,\\
\frac12(R_{uu}R_{vv})^{1/2}+\frac{E}{r^3}e^{-2f}+f_{uv}&\geq&0.
\end{eqnarray*}

Note how (as expected) some of these are the same as some of the
strong and weak energy conditions.

\subsection{Summary: covariant form of the energy conditions} The
energy conditions are given here in terms that use $r_{uv}$ and
$f_{uv}$. A more transparently covariant form is obtained by using
$R$ and $\Psi_2$ instead of these two. Then the results are as
follows (we note that the signs of $R_{uu}$, $R_{vv}$ and the term
$e^{4f}R_{uu}R_{vv}$ are invariants - the last of these is defined
in terms of contractions of Ricci with the two invariantly defined
radial null directions):

The strong energy condition is equivalent to
\begin{eqnarray}
R_{uu}&\geq& 0,\label{sec1}\\
R_{vv}&\geq&0,\label{sec2}\\
\frac12e^{2f}|R_{uu}R_{vv}|^{1/2}+2\frac{E}{r^3}+2\Psi_2-\frac{R}{12}&\geq&0,\label{sec3}\\
\frac12e^{2f}|R_{uu}R_{vv}|^{1/2}+\frac{E}{r^3}+\Psi_2-\frac{R}{6}&\geq&0.\label{sec4}
\end{eqnarray}

The weak energy condition is equivalent to (\ref{sec1}),
(\ref{sec2}), (\ref{sec3}) and
\begin{eqnarray}
\frac12e^{2f}|R_{uu}R_{vv}|^{1/2}+\frac{E}{r^3}+\Psi_2+\frac{R}{12}&\geq&0.\label{wec5}
\end{eqnarray}

The dominant energy condition is equivalent to (\ref{sec1}),
(\ref{sec2}), (\ref{sec3}) and
\begin{eqnarray}
\frac{E}{r^3}+\Psi_2+\frac{R}{12}&\geq&0,\label{dec6}\\
\frac12e^{2f}|R_{uu}R_{vv}|^{1/2}+\frac{R}{4}&\geq&0.\label{dec7}
\end{eqnarray}


\begin{thebibliography}{cauchy}
\bibitem[1]{carr} Carr B J and Coley A A {\em Class. Quantum
Grav.} {\bf 16} R31 (1999).
\bibitem[2]{OP} Ori A and Piran T {\em Phys. Rev.} {\bf D42}
1068 (1990).
\bibitem[3]{christo1} Christodoulou D {\em Ann. Math.} {\bf 140}
607 (1994).
\bibitem[4]{brady} Brady P R {\em Phys. Rev.} {\bf D51} 4168
(1995).
\bibitem[5]{bizon} Bizon P and Wasserman A {\em Class. Quantum Grav.} {\bf 19} 3309 (2002).
\bibitem[6]{gundlach} Gundlach, C {\em Living Rev. Rel.} {\bf 2} 4
(1999).
\bibitem[7]{harada} Harada T {\em Phys. Rev.} {\bf D63} 084022; {\em Class. Quantum Grav.} {\bf 18}
4549 (2001).
\bibitem[8]{christo2} Christodoulou D {\em Ann. Math.} {\bf 149}
183 (1999).
\bibitem[9]{nolan} Nolan B C {\em Class. Quantum Grav.} {\bf 18}
1651 (2001).
\bibitem[10]{brady2} Brady P R {\em Prog. Theor. Phys. Suppl.}
{\bf 136} 29 (1999).
\bibitem[11]{def1} We use the following standard notation for asymptotic
relations: $f(x)=O(g(x))$ as $x\to x_0$ iff $\exists$ constant $k$
s.t. $|f(x)|\leq k|g(x)|,\quad x\rightarrow x_0$. $f(x)=o(g(x))$
as $x\to x_0$ iff for any $\epsilon>0,|f(x)|\leq
\epsilon|g(x)|,\quad x\rightarrow x_0$. We define $\sim$ as $ f(x)
\sim g(x)$ as $x\to x_0$ iff $f(x)-g(x)=o(g(x))$ as $x\to x_0$.
\bibitem[12]{wald} Wald R M {\em General Relativity} (Univ.
Chicago Press, Chicago, 1984).
\bibitem[13]{foot} To use the method of Frobenius the
coefficients $q(x),p(x)$ should be analytic at $x=0$. However, to
obtain the required information about $H$ it is sufficient to use
a finite expansion with appropriate remainder terms, i.e.\ with
$q,p \in C^1$ at $x=0$. Thus we only require the metric
coefficients to be $C^2$ at $x=0$, and similarly at $x=x_c$. We
assume this henceforth.
\bibitem[14]{BendOr} Bender C M and Orszag S A {\em Asymptotic
Methods and Perturbation Theory} (Springer-Verlag, New York,
1999).
\bibitem[15]{debbie} Konkowski D A and Helliwell T
M
{\em Phys. Rev.} {\bf D54} 7898 (1996).
\end{thebibliography}
\end{document}